\documentclass[twocolumn,showpacs,floatfix,10pt,nofootinbib]{revtex4}
\usepackage{amssymb,amsmath}
\usepackage[dvips]{graphics,color}
\usepackage[colorlinks,hyperindex]{hyperref}
\usepackage{epsfig}\usepackage{float,stmaryrd,textcomp,bm,mathbbol}\usepackage{float,upgreek,yfonts}
\usepackage{graphicx}

\newcommand{\ba}{\begin{eqnarray}}
\newcommand{\ea}{\end{eqnarray}}
\newcommand{\bege}{\begin{equation}}
\newcommand{\enge}{\end{equation}}

\newcommand{\beq}{\begin{eqnarray}}
\newcommand{\benu}{\begin{enumerate}}
\newcommand{\enu}{\end{enumerate}}
\newcommand{\eeq}{\end{eqnarray}}

\hypersetup{hidelinks,backref=true,pagebackref=true,hyperindex=true,colorlinks=true,breaklinks=true,urlcolor= blue}
\hypersetup{%
  colorlinks = true,
  linkcolor  = blue
}

 \newcommand{\bea}{\begin{eqnarray}}
 \newcommand{\eea}{\end{eqnarray}}

\usepackage{hyperref}
\usepackage{xcolor}

\begin{document}

\title{Deformed AdS$_4$-Reissner-Nordstr\"om black branes and shear viscosity-to-entropy density ratio}

\author{A. J. Ferreira--Martins}
\email{andre.juan@ufabc.edu.br}
\affiliation{CCCCNH, Universidade Federal do ABC - UFABC,  09210-580, Santo Andr\'e, Brazil.}
\author{P. Meert}
\email{pedro.meert@ufabc.edu.br} 
\affiliation{CCCCNH, Universidade Federal do ABC - UFABC,  09210-580, Santo Andr\'e, Brazil.}
\author{R. da Rocha}
\email{roldao.rocha@ufabc.edu.br}
\affiliation{CMCC,  Federal University of ABC, 09210-580, Santo Andr\'e, Brazil.}


\begin{abstract}
A family of deformed AdS$_4$--Reissner--Nordstr\"om black branes, governed by a free parameter, is derived using the ADM formalism, in the context of the membrane paradigm. Their new event horizons, the Hawking temperature and other aspects are scrutinized. AdS/CFT near-horizon methods are then implemented to compute the shear viscosity-to-entropy ratio for the deformed AdS$_4$--Reissner--Nordstr\"om metric. The 
Killing equation is shown to yield new values for the free parameter and 
the shear viscosity-to-entropy ratio is used to derive a reliable range for  tidal charge. 
 \end{abstract}
\maketitle

\section{Introduction}
The AdS/CFT correspondence consists of a solid apparatus where 
 strongly coupled QFTs can be studied. Any given QFT, including 
 finite temperature ones, has a hydrodynamical description in the 
 infra-red (IR) limit, corresponding to long length scales in the theory. 
 In the anti-de Sitter (AdS)  bulk space, a theory of gravity is dual to the CFT on the boundary.  The AdS bulk geometry can include a black brane presenting an event horizon. The holographic duality between the AdS bulk and its boundary then conjectures that the CFT at the long
scale regime must be ruled by a near-horizon limit regarding 
dual geometries. For instance, any general relativistic  
black hole presents a spurious  fluid on its horizon, consisting of the so-called membrane paradigm, whose low-energy regime is a strongly coupled
field theory \cite{malda,Iqbal:2008by}. In the membrane paradigm,  black holes were scrutinized 
 \cite{Casadio:2015gea,daRocha:2017cxu}, in the long wavelength limit.  
Transport  coefficients were introduced by the duality between   black branes in the AdS bulk and fluid dynamics,  in the AdS boundary \cite{Bhattacharyya:2008jc}.  In the membrane paradigm, the AdS boundary can be identified with a brane, in the fluid/gravity  correspondence structure \cite{ssm1}. As a black brane 
is a solution of the Einstein's equations in the AdS bulk, the 4D brane 
can be also taken as an appropriate landscape, where  black holes on the brane are also solutions of the Einstein's effective field  equations. However, considering the 4D brane embedded in the AdS bulk, yields the AdS Riemann tensor to be related to the brane Riemann tensor by the Gauss--Codazzi equations. 
A useful constraint to the Einstein's effective field  equations on the brane \cite{ssm1} within the holographic membrane paradigm is to demand the general-relativistic limit, consisting of a rigid brane, meaning a brane that has infinite tension.  This condition, in fact,  produces a physically correct low-energy limit, allowing the construction of  black holes on the brane \cite{casadio,Ovalle:2017fgl,first,mgd1,mgd2}.

\par Methods in AdS/CFT and the membrane paradigm were successful derived in  Refs. \cite{Antoniadis:1998ig,Antoniadis:1990ew}. Tests of AdS/CFT
have been performed in cases for which the supergravity backgrounds are supersymmetric \cite{Meert:2018qzk,Bonora:2014dfa}. However, a
static AdS black hole in the supersymmetric limit can have a naked singularity, avoided when rotation is assumed, as in the Reissner--Nordstr\"om spacetime. Exploring the precise link between braneworld scenarios and the AdS/CFT duality, realized through the membrane paradigm \cite{ads_memb}, provides a transliteration of the brane models into the AdS/CFT language. It allows to scrutinize black branes and their hydrodynamics using the well-known AdS/CFT methods. In this work, we shall focus on computing transport coefficients, including the shear viscosity-to-entropy density ratio, $\eta/s$. Thereafter, viscosity bounds will be imposed to a generalized black brane.  
The Reissner--Nordstr\"om--AdS (AdS$_4$-RN) spacetime plays a prominent role in this procedure, being widely known in the context of AdS/CMT  \cite{Davison:2011uk,Cadoni:2009xm} for being dual to a finite temperature CFT describing a conserved $U(1)$ charge in the boundary. In fact there is a comprehensive literature regarding this spacetime and the related dual theory, regarding  holographic superconductors and strange metals \cite{Gubser:2008wv,Hartnoll:2008kx,Iqbal:2011in,Giordano:2018bsf}.

\par We will apply the ADM formalism to derive a family of spacetime  solutions, leading into the AdS$_4$-RN  in a very specific limit. One calls this spacetime a deformation of AdS$_4$-RN. It is governed by a free parameter in the radial component of the metric. When this parameter equals to unity, the AdS$_4$-RN  spacetime is then recovered. Although carrying similar features, the bulk geometry of the new solution is quite different of the standard AdS$_4$-RN  one. The first aspect noticed is the appearance of a coordinate singularity, which is in fact an event horizon when certain conditions are met. This opens new possibilities, as the horizon is an important feature when exploring the thermodynamic properties of a black hole. It has, thus, direct consequence in the CFT at the AdS boundary. Second, this spacetime can be derived from an action without matter terms, consisting of a vacuum in AdS/CFT, however
with a cosmological constant in an AdS$_5$, wherein AdS$_4$ is embedded. This is a consequence of the formalism implied to obtain the solution, and we use the notation $Q$ for the charge as mere analogy at this point, since its origin is different and the proper terminology is to call it a tidal charge \cite{maartens}, i. e., the source for this charge is the curvature of the spacetime itself, which comes from a higher dimensional AdS$_5$ bulk.

\par By allowing a free parameter in the  AdS$_4$-RN  metric,  we expect to gain more freedom, in the sense that intrinsic features, such as transport coefficients and thermodynamic quantities, can be modelled -- or fine tuned -- according to experimental evidence to come, or even eventually describe unknown materials in condensed matter.

\par This paper is organized as follows: in Sect. \ref{sec:2} the spacetime metric, that we call deformed AdS$_4$-RN , shall be derived  and some features considered relevant, such as the new horizon, its associated temperature and conditions for such quantities to be meaningful, are studied. We then proceed to describe the process of implementing perturbations to compute the shear viscosity-to-entropy ratio via the fluid/gravity correspondence in  Sect. \ref{sec:3}. In Sect.  \ref{sec:gkp_w} the AdS/CFT correspondence is briefly reviewed for our purposes and applied for computing the shear viscosity-to-entropy density ratio,  associated to the dual theory of the deformed AdS$_4$-RN  solution. Finally,  concluding remarks and some perspectives for the forthcoming developments regarding this type of solutions are presented in Sect. \ref{sec:5}. 

\section{Deformed AdS$_4$-Reissner-Nordstr\"om spacetime} \label{sec:2}

The  AdS$_4$--Reissner--Nordstr\"om black hole background represents a charged, asymptotically-AdS black hole solution with a planar horizon, described by coordinates $\{x^\mu\}=\{t,r,x,y\}$, 
\begin{eqnarray}\label{rnads4}
ds_4^2=-\frac{r^2 f(r)}{L^2}dt^2 +\frac{L^2}{r^2\,g(r)}dr^2+ \frac{r^2}{L^2}dx^2+ \frac{r^2}{L^2}dy^2,
\end{eqnarray}
with \begin{eqnarray}
f(r)=g(r)&=&1-(1+Q^2)\left(\frac{r_0}{r}\right)^3+Q^2\left(\frac{r_0}{r}\right)^4,\label{rnads41}
\end{eqnarray}
is a well known solution of the 4D Einstein--Maxwell theory with
a cosmological constant $\Lambda=-3/L^2$, where $Q$ denotes the black hole charge and $r_0$ represents the event horizon position  \cite{Davison:2011uk,Cadoni:2009xm}, which is unique, as the black hole is extremal. 

The AdS/CFT membrane paradigm can set the 4D brane as the boundary of an AdS$_5$ bulk with cosmological constant 
$\Lambda$ related to the vacuum energy density of the boundary.
The AdS$_5$ bulk, with electromagnetic field strength, satisfies the Einstein's equations, 
\begin{eqnarray}
{R}_{AB}-\frac12 R\,g_{AB}=\Lambda\,g_{AB}+{\color{black}{T_{AB}}},
\label{Dm1eq}
\end{eqnarray}
for 
\begin{equation}
{\color{black}{T_{AB}=F_{AC}F_{B}^{\text{ }C}-\frac{1}{4}g_{AB}F^{2},}}
\end{equation}
{\color{black}{where $F^{2}=F_{AB}F^{AB}$.}}
Projecting Eq. \eqref{Dm1eq} on an AdS$_4$ brane, with $\Lambda_4$ cosmological constant, and introducing Gaussian coordinates in the bulk, $x^\mu$ and $w$, one obtains the constraints at $w=0$:
\begin{eqnarray}
{R}_{\mu w}=0\ ,\ \ \
R=\Lambda_4 + E^\mu_{\;\mu},
\label{Deq}
\end{eqnarray}
where $R$ is the Ricci curvature scalar, $\Lambda_4$ the brane 
cosmological constant, and $E^\mu_\nu$ represents the electric part of the Weyl tensor. It mimics a Weyl fluid in the bulk \cite{maartens}, 
that illustrates how the brane embedding into the bulk contributes 
to the brane bending, possible by the finite value of the brane tension \cite{Ovalle:2017fgl,c_o_r_2014,Casadio:2016aum}.

For static solutions, Eqs.~(\ref{Deq}) emulate the Hamiltonian and the momentum  constraints \cite{Casadio:2001jg}, in the ADM  formalism. Such constraints play the important role of sorting out admissible field configurations along
manifolds of constant $w$ \cite{Casadio:2001jg}.
The selected field configurations thus can be propagated out the AdS$_4$ brane, under the residual Einstein's effective equations (\ref{Dm1eq}).
The Hamiltonian constraint comprehends  a
weaker stipulation than the 4D vacuum
equations $R_{\mu\nu}=0$. The $Q$ parameter is interpreted as a tidal charge \cite{maartens}.

Applying the ADM formalism to generate a family of deformed AdS$_4$--RN spacetimes, let us start with the metric \eqref{rnads4}, still considering the temporal components \eqref{rnads41},
\begin{eqnarray}\label{rngen}
ds_4^2=-\frac{r^2 f(r)}{L^2}dt^2 +n(r)\frac{L^2}{r^2}dr^2+ \frac{r^2}{L^2}dx^2+ \frac{r^2}{L^2}dy^2.
\end{eqnarray}
If instead one insists on requiring the AdS$_4$-RN  with a regular AdS$_4$ horizon, the price to pay is to have something beyond the vacuum in the bulk, as a field strength \cite{Davison:2011uk,Cadoni:2009xm}. This is not the case considered here. {\color{black}{ Considering that only the $rt$
component is non-zero, one can check that 
 $F^{2}=2F_{rt}F^{rt}$, and also obtain the components of 
the stress-energy tensor,  }}
\begin{subequations}
\begin{align}
{\color{black}{T_{tt}}} & {\color{black}{=F_{t}^{\text{ }r}F_{rt}-\frac{1}{4}g_{tt}F^{2}}}\\
{\color{black}{T_{rr}}} & {\color{black}{=F_{r}^{\text{ }t}F_{tr}-\frac{1}{4}g_{rr}F^{2}}}\\
{\color{black}{T_{xx}}} & {\color{black}{=T_{yy}=-\frac{1}{4}g_{xx}F^{2}}}
\end{align}
\end{subequations}
{\color{black}{Equivalently, the electromagnetic potential reads $A= A(r)\,dt$, having the chemical potential and the charge
density representing its components of expansion near boundary, $A(r)=\mu-\frac{{\rm Q}}{r}$. At the horizon, $r\to r_0$, $A\to 0$. Hence
$\mu=\frac{{\rm Q}}{r_0}$, in this regime. }}
{\color{black}{Here Q denotes the U(1) gauge charge, then $Q=\sqrt{{\rm Q}^2+\tilde{Q}^2}$, where $\tilde{Q}$ is the tidal charge}}.
\par A family of analytic solutions
of the form (\ref{rngen}) obtained by
relaxing the condition $f(r)=1/g(r)$ in (\ref{rnads4}), by fixing either $f$ or $g$ as
in the AdS$_4$-RN  and finding the most general solutions for
the constraints (\ref{Deq}).
These solutions will be expressed in terms of the ADM mass $M$
and the parameter $\beta$, which is fixed but arbitrary \cite{Abdalla:2009pg}.
The momentum constraints are identically satisfied by
the metric (\ref{rngen}) and the Hamiltonian constraint
can be written out explicitly in terms of $f(r)$ and $n(r)$.  
Fixing the temporal component (\ref{rnads41}), 
the Hamiltonian constraint in the ADM formalism reads
{\color{black}{{\begin{eqnarray}
-\!\!\!&\!\!\!&\!\!\!\!\frac{1}{324
   L^4 r^{14} p_Q^2 k_\beta^3}\nonumber\\&&
 \times \left\{22L^4 r^{12} \left(Q^2 r_0^4\!-\!r^4\!-\!r^2 r_0^2\!-\!r r_0^3+r^3r_0\right)\right.\nonumber\\&&\left.
   +4 L^4 r^{12}p_Q   \left(\left(Q^2\!+\!1\right) r r_0^3\!-\!3 Q^2 r_0^4\!-\!r^4\right)k_\beta^3\right.\nonumber\\
   &&\left.+4 L^4 r^8
   r_0 p_Q^2 k_\beta^2 
\left[9
   \left(Q^2\!+\!1\right) r^2 r_0^2+9 Q^2 r_0^4\!\right.\right.\nonumber\\&&\left.\left.-9 \left(2 Q^2\!+\!1\right) r r_0^3\!-\!2 (\beta \!-\!1)
   r^4\right]-L^4 r^{12} h_Q^2 k_\beta^3\!\right.\nonumber\\&&\left.
   +p_Q k_\beta^2 \left[9
   r^2-(\beta +17) r r_0+(2 \beta +7) r_0^2\right]\right.\nonumber\\&&\left.+81 r_0 (r-r_0)^2h_Q  p_Q^4
   \left[r^2 r_0^2 \left(2 \beta -27 Q^2-29\right)\right.\right.\nonumber\\&&\left.\left.+r r_0^3 \left[6 \beta +(4 \beta +59)
   Q^2+21\right]-4 (2 \beta +7) Q^2 r_0^4\right.\right.\nonumber\\&&\left.\left.+2 (\beta -1) (r^4+ r^3 r_0)\right]\right\}=0,\label{adm}
\end{eqnarray}}}}
for 
\begin{subequations}
\begin{eqnarray}
p_Q=p_Q(r)&=&\left(Q^2+1\right) r r_0^3-Q^2 r_0^4-r^4,\\
h_Q=h_Q(r)&=&p_Q-Q^2 r_0^4\\
k_\beta=k_\beta(r) &=& 9 r-(2 \beta
   +7) r_0,
   \end{eqnarray}
   \end{subequations}
   where $\beta$ denotes a free parameter that governs the deformation. It is free in the sense that its value is not determined \textit{a priori}, nevertheless it is taken as constant along the calculations. 
Hence, the solution 
\begin{equation}
	{\color{black}{ n(r)=\frac{1}{f\left(r\right)}\left\{ \frac{1-\frac{r_{0}}{r}}{1-\frac{r_{0}}{r}\left[1+\frac{1}{3}\left(\beta-1\right)\right]}\right\} }}\ , \label{eq:1}
\end{equation}
satisfies Eq. (\ref{adm}), once the temporal component is claimed to be the same as the standard AdS$_4$-RN, given by $f(r)$ in Eq. (\ref{rnads41}). 
Notice that for $\beta = 1$, the metric \eqref{rngen} is precisely the RN black hole in AdS$_4$ background. From now on $L=1$ will be assumed for the sake of conciseness.
{\color{black}{
\par The electromagnetic potential $A=A_{\mu}dx^{\mu}$ is determined from the Maxwell equations, which simply read 
\begin{equation}\label{eq:maxwell}
	\partial_{\mu}\left(\sqrt{-g}F^{\mu\nu}\right)=0\text{ },
\end{equation}
where $F_{\mu\nu}=\partial_{\mu}A_{\nu}-\partial_{\nu}A_{\mu}$ is the electromagnetic field strength. As assumed previously $A=A\left(r\right)dt$, the solution to \eqref{eq:maxwell} is given by 
\begin{widetext}
	\begin{equation} \label{eq:EMsol}
	A^{t}=\frac{-2Q\sqrt{r-r_{0}}\sqrt{\beta+2}\sqrt{3r-r_{0}\left(\beta+2\right)}-Qr\left(\beta-1\right)\left[\pi-2\arctan\left(\frac{\sqrt{r-r_{0}}\sqrt{\beta+2}}{\sqrt{3r-r_{0}\left(\beta+2\right)}}\right)\right]}{2\sqrt{3}r_{0}\sqrt{\beta+2}r}\text{ },
	\end{equation}
\end{widetext}
and one can verify that
\begin{equation}\label{eq:EMsolLimit}
	\lim_{\beta\rightarrow1}A^{t}=Q\left(\frac{1}{r}-\frac{1}{r_{0}}\right)\text{ },
\end{equation}
which is nothing but Coulomb's law - the solution for the spherically symmetric case, considered in the Reissner--Nordström scenario. Moreover one can check that
\begin{align}
	\lim_{r\rightarrow r_{\beta}}A^{t}&=0,\\
	\lim_{r\rightarrow\infty}A^{t}&\approx\mu-\frac{Q}{r}+\ldots \label{eq:asympSol}
\end{align}
It is of no surprise that these conditions are met, since those are the boundary conditions applied to \eqref{eq:maxwell}. In particular, expansion \eqref{eq:asympSol} at the boundary allows to obtain the chemical potential, $\mu$, of the CFT at the boundary. Explicitly, expanding \eqref{eq:EMsol} gives
\begin{equation}
	\mu=-\frac{Q}{6r_{0}}\left[6+\frac{\sqrt{3}\left(\beta-1\right)}{\sqrt{2+\beta}}\arctan\left(\sqrt{\frac{\beta+2}{3}}\right)\right]\text{ }.
\end{equation}

\par The essential features involving the event horizon of AdS$_4$-RN are present in the solution \eqref{rngen}. A closer look on \eqref{eq:1} reveals that $1/n(r_{\beta})\rightarrow 0$, i.e. $g^{rr}(r_{\beta})\rightarrow 0$, for
\begin{equation} \label{eq:4}
{\color{black}{r_{\beta}=r_{0}\left[1+\frac{1}{3}\left(\beta-1\right)\right]}}\,,
\end{equation}
Of course, this is a coordinate singularity at this point, but in fact one can ask weather it if $r_{\beta}$ is an event horizon or not. If so, by inspection, it is obvious that  $\beta >1$  leads to $r_{\beta}>r_0$, which means that \eqref{eq:4} is a bigger horizon than the standard AdS$_4$-RN outer horizon. Hence, it  should be considered as the surface for which quantities, such as temperature and entropy of the black hole, are evaluated.
\par
For a static space-time a sufficient condition for a surface to be an event horizon is that it be a Killing horizon \cite{wald}, which satisfies $\xi^{\mu}\xi_{\mu}=0$, where $\xi^{\mu}$ is the time-like Killing vector. Demanding that $\xi^{\mu}$ associated to \eqref{eq:1} satisfies such condition, two values of $\beta$ are found:
\begin{widetext}
	\begin{align}
	\beta&=1\ , \label{eq:6}								\\
	\color{black}\beta&\color{black}=-3+\frac{2\times2^{1/3}}{\left(-7-27Q^{2}+3\sqrt{3}\sqrt{3+14Q^{2}+27Q^{4}}\right)^{1/3}}+\frac{\left(7+27Q^{2}-3\sqrt{3}\sqrt{3+14Q^{2}+27Q^{4}}\right)^{1/3}}{2^{1/3}} \ . \label{eq:7}
	\end{align}
\end{widetext}
As expected, $\beta=1$ is a solution, which simply confirms that the  deformed AdS$_4$-RN contains the standard AdS$_4$-RN, for the previous established value of the deformation parameter. On the other hand, expression \eqref{eq:7} fixes $\beta$ according to the value of $Q$, {which is proportional to  the charge placed inside the event horizon when modelling the AdS$_4$-RN spacetime}. Eq. \eqref{eq:7} is not an intuitive function to picture, hence it is more useful to present its plot and point out certain features, useful for future arguments. Interestingly, it is seen that for both values, \eqref{eq:6} and \eqref{eq:7}, $r_{\beta}$ is a Killing horizon and, as was previously established, $\beta$ is a parameter that is not fixed \textit{a priori}. Moreover, Fig. 1 shows that the range $Q>\sqrt{3}$ implies $\beta > 1$, which is the region where \eqref{eq:4}, with $\beta$ given by \eqref{eq:7}, should replace $r_0$ for calculations regarding the horizon.}}
\begin{figure} [h!]
	\centering
	\includegraphics[scale=0.7]{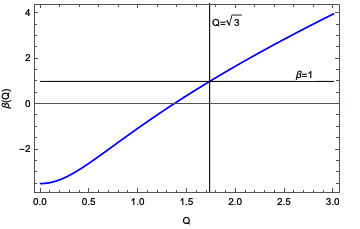}
	\label{fig:1}
	\caption{Plot of Eq. \eqref{eq:7}, $\beta(Q) \times Q$. It is  shown that $Q=\sqrt{3}$ yields $\beta=1$. Hence, we are interested in values of $Q>\sqrt{3}$ if considering $\beta$ given by Eq. \eqref{eq:7}.}
\end{figure}

\par The temperature of this black hole at the horizon $r_{\beta}$ is an important quantity, that will be computed according to the geometric procedure outlined in Ref. \cite{wald}. Such calculation proceeds as follows: one evaluates the quantity
\begin{equation} \label{eq:8}
	\kappa^{2}=\lim_{r\rightarrow r_{\beta}}\left[-\frac{\left(\xi^{\mu}\nabla_{\mu}\xi^{\lambda}\right)\left(\xi^{\nu}\nabla_{\nu}\xi_{\lambda}\right)}{\xi^{\rho}\xi_{\rho}}\right].
\end{equation} Then one identifies $\kappa$ as the surface gravity by arguing that the computation of such quantity leads to the same expression at the horizon. Then once one has the Killing vector, it is just a matter of computation. It is worth mentioning that Eq. \eqref{eq:8} is only valid at $r=r_{\beta}$, or at any other horizon if such surface exists. However, it cannot be used for other points in spacetime  \cite{wald}.  The surface gravity and temperature are closely related by 
\begin{equation} \label{eq:9}
	T=\frac{\kappa}{2\pi}\ .
\end{equation}
{\color{black}{Evaluating \eqref{eq:8} is fairly simple, and one finds immediately
\begin{align}
\begin{aligned}
\kappa^{2}&=\left.\frac{\left[2r^{4}+r_{0}^{3}\left(1+Q^{2}\right)r-2Q^{2}r_{0}^{4}\right]^{2}\left[3r+r_{0}\left(\beta+2\right)\right]}{12r^{4}\left(r_{0}-r\right)^{2}\left(r^{3}+r_{0}r^{2}+r_{0}^{2}r-r_{0}^{3}Q^{2}\right)}\right|_{r=r_{\beta}}\\&=0\label{tempads}
\end{aligned}
\end{align}
Hence, the temperature is identically zero. Notice that this result does not depend on the value of $\beta$, thus it is valid in the particular choice of $\beta$ given by \eqref{eq:7}. Moreover, we emphasize that this result is indeed expected, as we are considering an extremal black hole \cite{Zaanen:2015oix}.}}

\section{Fluid/gravity preliminaries} \label{sec:3}
\par This section is devoted to describe the theoretical background necessary to the computation of the shear viscosity-to-entropy density ratio, usually denoted by ${\eta}/{s}$, for the deformed RN-AdS$_4$ solutions.
\subsection{Linear response theory and Kubo's formula}
\label{sec:lrt}

\par Let one consider a theory described by an action $S$. It is often of interest to determine what is the response of a given operator $\mathcal{O}$, denoted by $\delta \langle \mathcal{O} \rangle$, when one adds a coupled external source, $\varphi^{(0)}$. Hence, the theory is altered as   $S \mapsto S + \int \mathrm{d}^4x \varphi^{(0)}(t, \bm{x}) \mathcal{O}(t, \bm{x})$ \cite{son_hydro}. The response is determined by employing the linear response theory, establishing that \begin{equation}
\delta \left \langle \mathcal{O} (t, \bm{x})\right \rangle := \left \langle \mathcal{O} (t, \bm{x})\right \rangle_S - \left \langle \mathcal{O} (t, \bm{x})\right \rangle \ ,
\end{equation}
where $\left \langle \mathcal{O} (t, \bm{x})\right \rangle_S$ denotes the ensemble average (one-point function) of the operator $\mathcal{O}$ in the presence of the external source.
The method is implemented by applying the time-dependent perturbation theory of quantum mechanics \cite{son_hydro,natsuume}, which yields the response, in momentum space,
\begin{equation}
\delta \left \langle \mathcal{O} (\omega, \bm{q})\right \rangle = - G_R^{\mathcal{O}, \mathcal{O}} (\omega, \bm{q}) \varphi^{(0)}(\omega, \bm{q}) \ ,
\label{eq:response_def}
\end{equation}
where $G_R^{\mathcal{O}, \mathcal{O}} (\omega, \bm{q})$ is the Fourier-transformed retarded Green's function associated to $\mathcal{O}$ \cite{natsuume}.

\par Eq. \eqref{eq:response_def} shows that the response to the operator, by a coupled external source, is narrowed down to the determination of the retarded Green's function, which is related to a transport coefficient through the Kubo's formula. Hereon we are ultimately interested in obtaining the shear viscosity $\eta$ as a transport coefficient emerging in the fluid dynamics that is dual to the deformed RN-AdS$_4$ family of solutions. 

\subsection{Hydrodynamics and fluid dynamics}
\label{sec:hydro}

\par The formalism of hydrodynamics provides a description of the macroscopic behaviour of a given system. More precisely, it refers to the dynamics of macroscopic variables, with main interest to the conserved ones, that remain in the hydrodynamic limit, characterized by a low-energy, long-wavelength regime \cite{natsuume}. Thus, from the field theoretical point of view, hydrodynamics is a legitimate  effective theory. Therefore, one cannot expect it to carry the details of a microscopic theory, which are encoded in the transport coefficients.

\par For a simple fluid, one can consider, as a macroscopic variable, the energy-momentum stress tensor, $T^{\mu \nu}$, along with its conservation law, $\partial_\mu T^{\mu \nu} = 0$. In $(3+1)$ dimensions, $T^{\mu \nu}$, a symmetric rank-$(2,0)$ tensor, has 10 independent components, whilst its conservation law provides only 4 equations. In order to close the equations of motion one must introduce a constitutive equation, which is conveniently written in terms of the normalized fluid velocity field $u^\mu(x^\nu)$, its rest-frame energy density $\rho(x^\mu)$ and its pressure $p(x^\mu)$.

\par In general, one introduces a constitutive equation by determining the form of $T^{\mu \nu}$ in a derivative expansion. To first order,  dissipation effects, which are absent in the perfect fluid -- which corresponds to the zeroth order -- are included. A viscous fluid has  stress tensor  expressed as \cite{Rangamani:2009xk}
\begin{equation}
T^{\mu \nu} = (\rho + p) u^\mu u^\nu + p \eta^{\mu \nu} + \tau^{\mu \nu} \ ,
\end{equation}
where the term $\tau^{\mu \nu}$ contains dissipative effects. In the local rest frame, it is such that\begin{equation}
\tau_{ij} = -\eta \left ( \partial_i u_j + \partial_j u_i -\frac{2}{3} \delta_{ij} \partial_k u^k \right ) - \zeta \delta_{ij} \partial_k u^k \ .
\end{equation}

\par Notice that two transport coefficients are introduced when dissipative effects are considered: the shear, $\eta$, and bulk, $\zeta$, viscosities. The introduction of a constitutive equation for the viscous fluid closes the equations of motion, yielding the continuity and the Navier--Stokes equations.

\subsection{Kubo's formula for the shear viscosity}
\label{sec:kubo_visco}

\par General relativity states that spacetime fluctuations bring up fluctuations into the stress tensor \cite{wald}. In agreement with this idea, the Kubo's formula for the shear viscosity $\eta$ can be derived by coupling fictitious gravity to the fluid, and then determining the response of $T^{\mu \nu}$, under gravitational perturbations. At this stage, one should see this procedure just as a quick way to derive the Kubo's formula, as one does not really curve spacetime in fluid experiments. However, the derivation presented here has a natural interpretation within the AdS/CFT duality framework, as discussed in Sec. \ref{sec:gkp_w}.
\par First, one adds a gravitational perturbation to a 4D  spacetime. Since we are interested in determining the shear viscosity $\eta$, it is natural to consider an off-diagonal perturbation, so that the perturbed metric $g_{\mu \nu}^{(0)}$ is, in the $\{t, x, y, z\}$ coordinate system, given by 
\begin{equation}
g_{\mu \nu}^{(0)}\mathrm{d}x^\mu \mathrm{d}x^\nu = \eta_{\mu \nu} \mathrm{d}x^\mu \mathrm{d}x^\nu + 2h_{xy}^{(0)}(t) \mathrm{d}x \mathrm{d}y \ ,
\label{eq:perturbed_metric}
\end{equation} for $\eta_{\mu \nu}$ 

\par Since the perturbed spacetime is no longer flat, one must extend the constitutive equation for the stress tensor to a curved spacetime, according to
\begin{equation}
T^{\mu \nu} = (\rho + p) u^\mu u^\nu + P g^{\mu \nu(0)} + \tau^{\mu \nu} \ ,
\end{equation}
so that the dissipative part now reads 
\begin{equation}
\begin{gathered}
\tau^{\mu \nu} = - \eta P^{\mu \sigma} P^{\nu \lambda} \left ( \nabla_\sigma u_\lambda + \nabla_\lambda u_\sigma - \frac{2}{3} g_{\sigma \lambda}^{(0)} \nabla_k u^k \right )  + 
\\ 
-\zeta P^{\mu \sigma} P^{\nu \lambda} g_{\sigma \lambda}^{(0)}  \nabla_k u^k  \ ,
\label{eq:tau_covar}
\end{gathered}
\end{equation}
where $\nabla_\mu$ represents the covariant derivative with respect to the perturbed metric $g_{\mu \nu}^{(0)}$. The $P^{\mu \nu} := g^{\mu \nu (0)} + u^\mu u^\nu$ is the projection tensor, necessary to write the constitutive equation in a covariant way.

\par Notice that the perturbation is considered to be homogeneous, as well as the fluid velocity field, i.e., $u_i = u_i(t)$. However, parity invariance forbids motion in any direction, so the fluid must be at rest. 
Therefore, the covariant derivative of the velocity field is reads  $\nabla_\mu u_\nu = \partial_\mu u_\nu - \Gamma^\sigma_{\mu \nu} u_\sigma = \Gamma^t_{\mu \nu}$. Calculating the response in $\tau^{xy}$, up to linear order in the perturbation, requires the covariant derivatives $\nabla_x u_y = \Gamma^t_{xy} = \Gamma^t_{yx} = \nabla_y u_x$. Notice that they are of linear order in the perturbation, 
\begin{equation}
\begin{gathered}
\!\!\!\!\!\!\Gamma^t_{xy} \!=\! \frac{1}{2} g^{tt (0)} \left ( \partial_x g_{ty}^{(0)} \!+\! \partial_y g_{tx}^{(0)} \!-\! \partial_t g_{xy}^{(0)} \right ) \!=\! \frac{1}{2} \partial_t h_{xy}^{(0)} \ , 
\end{gathered}
\end{equation}
so that the terms proportional to $ \nabla_k u^k$ in Eq. \eqref{eq:tau_covar} will be second order in the perturbation. Now, since $ \nabla_x u_y$ is already linear in the perturbation, one can use the projection tensor in the rest frame and in flat spacetime, $P^{\mu \nu} = \text{diag}(0, 1, 1, 1)$. In fact, considering the perturbed metric contribution yields in terms an order higher in the perturbation. Therefore, the response in $\tau^{xy}$ reads
\begin{equation}
\delta \left \langle \tau^{xy} \right \rangle = -2 \eta \Gamma^t_{xy} = -\eta \partial_t h_{xy}^{(0)} \ ,
\end{equation} 
whose Fourier transformation yields 
\begin{equation}
\delta \left \langle \tau^{xy} (\omega, \bm{0}) \right \rangle = i \omega \eta h_{xy}^{(0)} \ .\label{oooo}
\end{equation}
Comparing Eq. (\ref{oooo}) with the general result from linear response theory of Eq. \eqref{eq:response_def}, in this case, $\delta \left \langle \tau^{xy} \right \rangle = - G_R^{xy, xy} h_{xy}^{(0)}$, one obtains the Kubo's formula, 
\begin{equation}
\eta = - \lim_{\omega \rightarrow 0} \frac{1}{\omega} \Im \left ( G_R^{xy, xy} (\omega, \bm{0}) \right )\ .
\end{equation}
Naturally, $\eta$ does depend neither on $\omega$ nor on $\bm{q}$, which is why one takes the $\omega \rightarrow 0$ limit and set $\bm{q} = \bm{0}$ in the Green function, accounting to $k = 0$. Of course, the problem of finding the retarded Green's function $G_R^{xy, xy}$ remains, and one will  employ the AdS/CFT methods to this task.

\section{AdS/CFT and the GKP--Witten relation}
\label{sec:gkp_w}

\par The core claim of the AdS/CFT duality \cite{malda} is that the generating functionals, or partition functions, of the gauge and gravitational theories are equivalent, $Z_{gauge} = Z_{AdS}$, which is realized through the (Lorentzian) GKP--Witten relation \cite{gkp1,gkp2}:
\begin{equation}
\left \langle \exp \left ( i \int \varphi^{(0)} \mathcal{O} \right )\right \rangle = \exp\left (i \bar{S}[\varphi^{(0)}]\right ) \ ,
\label{eq:gkp_witten}
\end{equation}
where $\langle \;\cdot\; \rangle$ denotes an ensemble average. The  $\varphi$ represent a field in the gravitational (bulk) theory, and $\bar{S}$ is the on-shell action. Also, $\varphi^{(0)} = \left. \varphi \right \rvert_{u=0}$, in coordinates such that the AdS boundary where the gauge theory lives is located at $u=0$ -- for the spacetime we are working with, c.f. Eq. \eqref{rngen}, one has $u=r_{0}/r$. Hereon the  coordinate $u$ will be used instead of $r$, that is, the functions appearing in the metric coefficients  \eqref{rnads41} and \eqref{eq:1} are now $f=f(u)$ and $n=n(u)$, respectively.

\par The left-hand side of Eq. \eqref{eq:gkp_witten} is the generating functional of the $D$-dimensional boundary gauge theory, when an external source $\varphi^{(0)}$ is added, whilst the right-hand side of Eq. \eqref{eq:gkp_witten} is the generating functional of a $(D+1)$-dimensional bulk gravitational theory. The on-shell action is obtained by simply evaluating the integral when the field $\varphi$ is the solution of the equations of motion meeting certain conditions imposed at the AdS boundary, $\left .\varphi \right \rvert_{u=0} = \varphi^{(0)}$. Now, since $\varphi$ is the solution of the equation of motion, $\bar{S}$ reduces to a surface term on the AdS boundary, which allows us to obtain from the $(D+1)$-dimensional action a $D$-dimensional quantity, which is identified with the generating functional of the boundary theory, according to Eq. \eqref{eq:gkp_witten}. This is the sense in which we loosely say that the gauge theory lives on the boundary of the bulk.

\par An important point to notice is that, from the $(D+1)$-dimensional point of view, $\varphi$ is a field propagating in the bulk; whilst $\varphi^{(0)}$ is an external source from the $D$-dimensional point of view. Therefore, in the context of AdS/CFT, one can say that a bulk field acts as an external source of a boundary operator.

\par In practice, the greatest operational advantage provided by the GKP--Witten relation is the possibility of obtaining the generating functional of a gauge theory by the evaluation of the classical action of a gravitational theory. As we are interested in the response of a system in the presence of an external source, one obtains, from the GKP--Witten relation, the following expression for the one-point function \cite{natsuume}, 
\begin{equation}
\left \langle \mathcal{O} \right \rangle_S = \frac{\delta \bar{S}[\varphi^{(0)}]}{\delta \varphi^{(0)}} \ .
\label{eq:one_point}
\end{equation}
Naturally, to obtain the one-point function in the absence of the external source, one evaluates the expression above for $\varphi^{(0)} = 0$, that is, $\left \langle \mathcal{O} \right \rangle = \left . \left \langle \mathcal{O} \right \rangle_S \right \rvert_{\varphi^{(0)} = 0}$.

\par The gravitational theory to be considered is classical 4D  general relativity with negative cosmological constant, $\Lambda<0$, i.e., the Einstein--Hilbert action plus a matter term
\begin{equation}\label{eq:fullAction}
	S=\frac{1}{16\pi}\int d^{4}x\sqrt{-g}\left(R-2\Lambda\right)+S_{mat} \ ,
\end{equation}
where $S_{mat}$ is chosen according to the boundary theory. {\color{black}{Besides, as the term $F^2$ can be included into the part $S_{mat}$, since it does not contribute for the computation of the shear viscosity, $\eta$. Although it does contribute to the computation of other transport coefficients, as the conductivities, which is not our aim here}}. 

The Here a massless scalar field is considered
\begin{equation} \label{eq:scalarAction}
	S_{mat}=\int d^{4}x\sqrt{-g}\left(-\frac{1}{2}\left(\nabla_{\mu}\varphi\right)^{2}\right)\ .
\end{equation}
In fact, since one is only interested in the asymptotic behaviour, the metric \eqref{rngen} reads 
\begin{equation} \label{eq:assympMetric}
	g_{\mu\nu}\mathrm{d}x^{\mu}\mathrm{d}x^{\nu}\sim\frac{r_{0}^{2}}{u^{2}}\left(-\mathrm{d}t^{2}+\frac{1}{r_{0}^{2}}\mathrm{d}u^{2}+\mathrm{d}x^{2}+\mathrm{d}y^{2}\right)\ ,
\end{equation}
which is, of course, the AdS spacetime. In these coordinates the equation of motion from Eq. \eqref{eq:scalarAction} is
\begin{equation} \label{eq:scalarEOM}
	\left(\frac{1}{2u^{3}}\varphi^{\prime}\right)^{\prime}\sim0\ ,
\end{equation}
where the prime denotes differentiation with respect to $u$. Notice that the Einstein--Hilbert action does not play any role in the calculation of $\left\langle \mathcal{O}\right\rangle _{S}$, since it is independent of the field $\varphi$. Requiring the scalar field to be static and homogeneous along the boundary direction, $\varphi=\varphi\left(u\right)$, yields 
\begin{eqnarray} \label{eq:assympAction}
	\!S_{mat}\!&=&\!\int d^{4}x\sqrt{-g}\!\left(\!-\frac{1}{2}g^{uu}\left(\!\varphi^{\prime}\!\right)^{2}\!\right)\nonumber\\&&\sim\int dx^{4}\left(-\frac{1}{2}\frac{r_{0}^{3}}{u^{2}}\left(\varphi^{\prime}\right)^{2}\right)\ .
\end{eqnarray}
The usual technique to proceed is integration by parts, leading to \begin{equation} \label{eq:actionParts}
	\!\!\!S_{mat}\!\sim\!\!\int\! d^{3}x\!\left.\left(\frac{r_{0}^{3}}{2u^{2}}\varphi\varphi^{\prime}\right)\right|_{u=0}\!\!\!+\!\int d^{4}xr_{0}^{3}\!\left(\frac{1}{2u^{3}}\varphi^{\prime}\right)^{\prime}\!\varphi \ ,
\end{equation}
where it is assumed that the scalar field vanishes at the horizon. The second term in Eq. \eqref{eq:actionParts} is just the equation of motion \eqref{eq:scalarEOM}, which has the following asymptotic solution
\begin{equation} \label{eq:scalarSol}
	\varphi\sim\varphi^{\left(0\right)}\left(1+\varphi^{\left(1\right)}u^{3}\right) \ .
\end{equation}
For a scalar field of this form, Eq. \eqref{eq:scalarSol}, the action \eqref{eq:actionParts} reduces to a surface term on the AdS boundary. Substituting \eqref{eq:scalarSol} on the action and evaluating the surface term, the on-shell action reads 
\begin{equation} \label{eq:actionOnshell}
	\bar{S}\left[\varphi^{\left(0\right)}\right]\sim\int d^{3}x\left(\frac{3}{2}r_{0}^{3}\left(\varphi^{\left(0\right)}\right)^{2}\varphi^{\left(1\right)}\right) \ .
\end{equation}
Thus, the one-point function from \eqref{eq:one_point} is given by 
\begin{equation} \label{scalarOnepoint}
	\left\langle \mathcal{O}\right\rangle _{S}=3r_{0}^{3}\varphi^{\left(1\right)}\varphi^{\left(0\right)} \ .
\end{equation}
which, compared to the linear response relation of Eq. \eqref{eq:response_def}, yields the Green function:
\begin{equation}
G_R^{\mathcal{O}, \mathcal{O}} (k = 0) = -3 r_0^3 \varphi^{(1)} \ ,
\end{equation}

\subsection{$\eta/s$ in the deformed RN-AdS$_4$ black brane}
\par A bulk perturbation $h_{xy}$ can be considered, such that
\begin{equation}
ds^2 = ds^2_{0} + 2h_{xy} \mathrm{d}x \mathrm{d}y \ .
\end{equation}
where $ds^{2}_{0}$ is given by \eqref{rngen}. According to the linear response theory, the response in the boundary energy-momentum tensor is given by 
\begin{equation}
\delta \left \langle \tau^{xy} \right \rangle  = i \omega \eta h_{xy}^{(0)} \ ,
\label{eq:response_1_2}
\end{equation}
where $h_{xy}^{(0)}$ is the perturbation added to the boundary theory, which is asymptotically related to $h_{xy}$ by
\begin{equation}
g^{xx}h_{xy} \sim h_{xy}^{(0)} \left ( 1 + h_{xy}^{(1)} u^3 \right ) \ ,
\label{eq:perturb_asym_def_2}
\end{equation}
holding as long as $g^{xx}h_{xy}$ obeys the equation of motion for a massless scalar field \cite{natsuume, son_hydro}. Since the asymptotic behaviour of \eqref{rngen} is given by \eqref{eq:assympMetric}, one  can directly apply the results discussed in this section, and treat $g^{xx}h_{xy}$ as the 4D bulk scalar field $\varphi$, which acts as an external source of a boundary operator. Thus, Eq. \eqref{scalarOnepoint} yields 
\begin{equation}
\delta \left \langle \tau^{xy} \right \rangle = \frac{r_0^3}{16 \pi}3 h_{xy}^{(1)}h_{xy}^{(0)} \ .
\label{eq:response_2_2}
\end{equation}
Comparing Eqs. \eqref{eq:response_1_2} and \eqref{eq:response_2_2} implies that 
\begin{equation}
i \omega \eta = \frac{3r_0^3}{16\pi G_4} h_{xy}^{(1)} \ .
\label{eq:eta_quase_2}
\end{equation}

\par On the other hand, the entropy density associated to \eqref{rngen} is, according to the area law, given by $s = r_{\beta}^2/4$\footnote{Recall that we are concerned about the horizon given by $r_{\beta}$, c.f. Eq. \eqref{eq:4}, with $\beta$ given by \eqref{eq:7}.}. Plugging this result into Eq. \eqref{eq:eta_quase_2} yields
\begin{equation}
\begin{gathered}
\frac{\eta}{s} = \frac{3r_0}{4\pi\left[1+\frac{2}{9}\left(\beta-1\right)\right]^{2}} \frac{h_{xy}^{(1)}}{i \omega} \ .
\label{eq:eta_s_geral_2}
\end{gathered}
\end{equation}

\par One now must find $h_{xy}^{(1)}$, by solving the equation of motion for the perturbation $g^{xx}h_{xy} \equiv \varphi$, which is that of a massless scalar in a 4D background, 
\begin{equation}
\partial_\mu \left ( \sqrt{-g} g^{\mu \nu} \partial_\nu \varphi \right ) = 0
\end{equation}
Taking $\varphi = \phi(u) e^{-i\omega t}$, the perturbation equation reduces to
\begin{equation}
\frac{u^2}{\sqrt{fn}} \left ( \frac{\sqrt{fn}}{u^2}\phi^{\prime}\right )^{\prime}  + \frac{1}{fn} \frac{\omega^2}{r_0^2} \phi  = 0 \ .
\label{eq:pertu_edo_2}
\end{equation}
Now, by imposing the incoming wave boundary condition near the horizon and the Dirichlet boundary condition at the AdS boundary, {and proceeding as outlined in \cite{natsuume} to incorporate these conditions}\footnote{Namely: solving Eq. \eqref{eq:pertu_edo_2} in the limit $u \rightarrow 1$ and afterwards in a power series of $\omega$ up to $\mathcal{O}(\omega)$ in the hydrodynamic limit.} one arrives at the full solution 
\begin{equation}
\phi = \phi^{(0)} \left ( 1 - i \frac{\omega}{r_0} \frac{Q^2-3}{|Q^2-3|} \int \frac{u^2}{\sqrt{fn}}\, \mathrm{d} u \right ) \ .
\end{equation}
\par Accordingly, the full time-dependent perturbation is asymptotically given by
\begin{equation}
g^{xx}h_{xy} \sim e^{-i\omega t} \phi^{(0)}  \left ( 1 - i \frac{\omega}{r_0} \frac{Q^2-3}{|Q^2-3|}\frac{u^3}{3} \right )  \ .
\label{eq:perturb_asym_solution_2}
\end{equation}
Comparing now Eq. \eqref{eq:perturb_asym_solution_2} to Eq. \eqref{eq:perturb_asym_def_2}, and identifying $h_{xy}^{(0)} =  \phi^{(0)} e^{-i \omega t}$, one promptly obtains 
\begin{equation}
h_{xy}^{(1)} = \frac{-i\omega}{3r_0}  \frac{Q^2-3}{|Q^2-3|} \ .
\end{equation}
\noindent Thus, substituting the resulting $h_{xy}^{(1)}$ above in Eq. \eqref{eq:eta_s_geral_2} yields
	\begin{equation} \label{eta-sRatio}
		\frac{\eta}{s}=\color{black}\frac{9}{\left(1-\frac{2\times2^{1/3}}{\chi_{Q}}+\frac{\chi_{Q}}{2^{1/3}}\right)^{2}} \left(\frac{3-Q^{2}}{\left|3-Q^{2}\right|}\right)\ ,
	\end{equation}
	for $\color{black}\chi_Q=\left(-7-27Q^{2}+3\sqrt{9+42Q^{2}+81Q^{4}}\right)^{1/3}$, 
where the $\beta$ parameter is written as given by Eq. \eqref{eq:7}.

\par
\begin{figure} [H]
	\centering
	\includegraphics[scale=0.7]{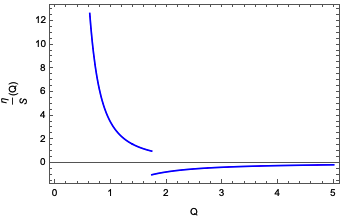}
	\caption[short text]{Plot of Eq. \eqref{eta-sRatio} without the factor $\frac{1}{4\pi}$. We see that there is a change of sign in the $\eta/s$ ratio for $Q=\sqrt{3}$.}	\label{eta-s-plot1}
\end{figure}

 The ratio $\eta/s$ is a positive quantity, since both the shear viscosity and the entropy density are positive. From the plot in Fig.  \ref{eta-s-plot1}, representing Eq. \eqref{eta-sRatio}, one sees where this condition is met. 
Therefore, the following bound for $Q$ can be obtained
\begin{equation}
\begin{gathered}
0 < Q < \sqrt{3} \ , 
\end{gathered}
\end{equation}
for the $\eta/s$ ratio to assume the saturated value $\eta/s = 1/4\pi$ \cite{kss}, as the initial action is Einstein--Hilbert. Notice that, as defined, the tidal charge $Q$ must be a positive quantity, so that the $-\sqrt{3} < Q < 0$ interval, which also satisfies the $3-Q^2 > 0$ bound, was not considered in this result, which is also very interesting\textcolor{black}{, and worthy of further investigation}.

\section{Concluding remarks and perspectives} \label{sec:5}

The membrane paradigm was here used to derive a new 
 family of deformed AdS$_4$--RN black branes.
 For it, the ADM formalism was employed, assuming the temporal component fixed, and the Hamiltonian constraint implemented a deformation of the  AdS$_4$--RN black brane, given by Eqs. (\ref{rngen}, \ref{eq:1}). The Killing equation 
 for the Killing vector of the deformed AdS$_4$-RN black brane 
 was then solved, yielding the values of the $\beta$ free parameter given by Eqs. (\ref{eq:6}) and (\ref{eq:7}). Although Eq. (\ref{eq:6})
 yields the standard AdS$_4$-RN black brane, Eq. (\ref{eq:7}) presents new solutions, encoding the deformed AdS$_4$-RN black brane $\beta$ free parameter a function of the black brane tidal charge. The Hawking temperature was also computed in Eq. (\ref{tempads}). 
 Fluid/gravity methods were then employed for computing the shear viscosity-to-entropy ratio for the deformed AdS$_4$--Reissner--Nordstr\"om black branes. This ratio is used to derive a reliable range for the  tidal charge as well.

It is worth to emphasize that the value of the $\beta$ deformation parameter, in Eq. (\ref{eq:6}), does correspond to the standard AdS$_4$--RN black brane, as expected. 
 In fact by exploring some features of the deformation exposed in Sect. \ref{sec:2}, we found that the deformation parameter is restricted to a precise value, taking us back to the conventional AdS$_4$-RN spacetime, which we see as an argument in favour of the unicity of this solution, whenever the ADM formalism is utilized.

Preliminary numerical analysis provide us with  deformations of the AdS$_4$--RN black brane (\ref{rnads4}, \ref{rnads41}) without AdS$_5$ bulk embedding, as an exact solution of a Lee--Wick-like action of gravity, and then more members of the family of deformed 
AdS$_4$-RN  black branes might be taken into account, with free parameter given by Eq. (\ref{eq:7}). In order to implement it, the value of $\eta/s$ cannot be conjectured to be equal to $1/4\pi$, and must be derived for the Lee--Wick-like action. However, up to now we have  not concluded these computations, as the machining time employed is awkwardly long. If these calculations can be finally implemented, and if the $\eta/s$ ratio 
allows a value $\beta\neq 1$, one can therefore apply the deformed AdS$_4$-RN black brane in the context of the AdS/CMT correspondence. In fact, the standard AdS$_4$--RN black brane is already known to describe the  strange metals in the holographic duality setup. By promoting the spacetime metric from the standard AdS$_4$--RN black brane, corresponding to the particular value in Eq. (\ref{eq:6}),  to the deformed family (\ref{rnads41}, \ref{rngen}) studied here, with the $\beta$ parameter given by Eq. (\ref{eq:7}), we expect to gain freedom in describing such materials. Hence this family of deformed black branes can  model a wider class of strange metals and the $\beta$ parameter in Eq. (\ref{eq:7}) can be then used to compute other transport coefficients, or fine tune quantities already known, like
the electric conductivity and the thermal conductivity, for example.

\section*{Acknowledgements}
\textcolor{black}{AJFM} is grateful to FAPESP (Grants No.  2017/13046-0  and No.  2018/00570-5) and to Coordena\c c\~ao de Aperfei\c coamento de Pessoal de N\'ivel Superior - Brazil. The work of PM was financed by CAPES. 
 RdR~is grateful to CNPq (Grant No. 303293/2015-2),  to FAPESP (Grant No. 2017/18897-8), and to ICTP HE 210-V, for partial financial support. 

\bibliography{bibliografia}
\bibliographystyle{iopart-num}

\end{document}